\newcommand{\CLASSINPUTbaselinestretch}{1.6}
\documentclass[journal,12pt,onecolumn,draftclsnofoot,]{IEEEtran}
\IEEEoverridecommandlockouts
\usepackage{cite}
\usepackage{amsmath,amssymb,amsfonts}
\usepackage{algorithmicx}
\usepackage{graphicx}
\usepackage{textcomp}
\usepackage{subcaption}
\usepackage{subfloat}
\usepackage{xcolor}
\usepackage{algpseudocode}
\usepackage[ruled,vlined]{algorithm2e}
\def\BibTeX{{\rm B\kern-.05em{\sc i\kern-.025em b}\kern-.08em
    T\kern-.1667em\lower.7ex\hbox{E}\kern-.125emX}}

\begin{document}
\title{Deep Learning Based Load Balancing for improved QoS towards 6G}
\DeclareRobustCommand*{\IEEEauthorrefmark}[1]{%
  \raisebox{0pt}[0pt][0pt]{\textsuperscript{\footnotesize #1}}%
}

\author{\IEEEauthorblockN{Vishnu Vardhan Nimmalapudi\IEEEauthorrefmark{1},
Ajith Kumar Mengani\IEEEauthorrefmark{1},
Roopa Vuppula\IEEEauthorrefmark{2}, 
Rahul Jashvantbhai Pandya\IEEEauthorrefmark{3}}
\IEEEauthorblockA{\IEEEauthorrefmark{1,}\IEEEauthorrefmark{2,}\IEEEauthorrefmark{3} Department of Electronics and Communication Engineeering,
NIT Warangal,
India\\
Email: \IEEEauthorrefmark{1}nvishnu@student.nitw.ac.in,
\IEEEauthorrefmark{1}majith@student.nitw.ac.in,
\IEEEauthorrefmark{2}roopav44@student.nitw.ac.in,
\IEEEauthorrefmark{3}rpandya@nitw.ac.in,
}}

\maketitle
\begin{abstract}
Deep learning has made great strides lately with the availability of powerful computing machines and the advent of user-friendly programming environments. It is anticipated that the deep learning algorithms will entirely provision the majority of operations in 6G. One such environment where deep learning can be the right solution is load balancing in future 6G intelligent wireless networks. Load balancing presents an efficient, cost-effective method to improve the data process capability, throughput, and expand the bandwidth, thus enhancing the adaptability and availability of networks. Hence a load balancing algorithm based on Long Short Term Memory (LSTM) deep neural network is proposed through which the base station’s coverage area changes according to geographic traffic distribution, catering the requirement for future generation 6G heterogeneous network. The LSTM model’s performance is evaluated by considering three different scenarios, and the results were presented. Load variance coefficient (LVC) and load factor (LF) are introduced and validated over two wireless network layouts (WNL) to study the Quality of Service (QoS) and load distribution. The proposed method shows a decrease of LVC by 98.311\% and 99.21\% for WNL1, WNL2 respectively.
\end{abstract}


\section{Introduction}
The fifth-generation (5G) of wireless technology was commercially deployed across 34 countries as of Jan. 2020. Hence research has begun to examine beyond 5G and to gestate the sixth-generation (6G). Deep Learning algorithms had shown promising results in the past few years in the field of communication systems\cite{b1,b2}, motivating future 6G wireless networks to make use of these algorithms. The artiﬁcial intelligence (AI) enabled 6G intelligent  network\cite{b3} is expected to be deployed between 2027 and 2030.  The 6G wireless system needs to be introduced with new captivating features by using new technologies and simultaneously continuing the trends of the previous generations. The most powerful technologies and services like unmanned aerial vehicles (UAV), AI,  3D networking, autonomous vehicles, smart wearables, terahertz (THz) band, implants,  sensing, 3D mapping, optical wireless communication (OWC), wireless power transfer, and computing reality devices, will be the driving force for 6G \cite{b4}. With the rapid growth of various emerging applications, such as virtual reality, Internet of Everything (IoE), and the three-dimensional (3D) media that require high data rates, the design goals for 6G, its implementation strategies, and challenges are already being explored in literature\cite{b5,b6,b7,b8}. 5G can reach up to  20 Gb/s per-user bit rate for end-users with millimeter-wave (mm-wave) communications and large-scale antenna arrays, i.e., massive multiple-input and multiple-output (MIMO). Meanwhile, it is envisioned that the terahertz (THz) band in 6G will serve as the next frontier for communications with a per-user bit rate of approximately  1Tb/s in most cases\cite{b9}. Hence, the capacity to handle huge volumes of data and provide high-data-rate connectivity per device is the essential requirement for 6G intelligent networks. The load balancing technique can serve as a promising solution to efficiently handle higher data rates and manage wireless resource allocation among multiple connections. It also offers improved system performance, higher resource utilization, and decreased operational cost. 
 
There are several methods in the literature for load balancing.
 S. T. Girma et al. proposed a fuzzy logic-based load balancing algorithm\cite{b10}. In this system, the traffic load is balanced by transferring some of the ongoing calls of densely loaded Base Transceiver Stations (BTS) to the underloaded BTS. The handoff index is calculated for the serving BTS and all its neighboring BTS, and once the neighboring BTS is identified better, the handoff process is executed.  The minimum value of this fuzzy system's output is 0, which means no handoff, and the maximum value is 1, indicating exactly handoff. Y.  Bejerano presented a load balancing scheme based on controlling the size of Wireless Local Area Network (WLAN) cells as like cell breathing\cite{b11}. It uses two algorithms: first is to decrease the load of congested AP(s), and the second is to provide an optimal min-max load-balanced solution. While J. Wu compare coverage shaping and Adaptive tilting based load balancing techniques \cite{b12}. It is proved that the coverage shaping based approach gives the best performance enhancement compared to the adaptive tilting system since the Bubble Oscillation Algorithm (BOA)\cite{b13} in coverage shaping technique finds the optimal boundaries to serve the massive demand. R. Misra et al. presented the machine learning oriented dynamic cost factors based routing in communication networks\cite{b14}.
 
The above-mentioned load balancing techniques did not consider traffic priority as a criterion for balancing the load. Instead, they have only considered the traffic density. They are based on balancing the load by changing the coverage area for every few seconds, which may result in some power loss. Hence there is a trade-off between the power loss due to frequent change in coverage areas and the efficiency of load balancing. Therefore a predetermined load balancing technique is required through which the coverage area can be changed only a few times without changing the load balancing efficiency by much. Hence we propose a load balancing technique based on prioritized traffic prediction through which the area that should be covered by the base stations is predicted one day ahead. The load among the base stations is divided almost equally. The network traffic is prioritized into three default Quality of Service (QoS) Classes: High Priority (Priority-1 or P1), Medium Priority (Priority-2 or P2), and Low Priority (Priority-3 or P3). The traffic, which is essential or time-critical and needs to be processed without any delay, is considered a high priority. Generally, military communications, banking transactions, and videos on Amazon Prime, Netflix, YouTube, video games live-streaming sites are considered as high priority, whereas emails, Facebook, Whatsapp are considered as Medium Priority, and unessential browsing data is considered as Low Priority.
 
There are several methods available in the literature for wireless traffic modeling and prediction. Y.  Shu et al. applied the most widely used time series forecasting method, Auto-Regressive Integrated Moving Average (ARIMA), for traffic load prediction \cite{b15}. S. Jaffry proved that the Long short-term memory (LSTM) and vanilla feed-forward neural networks (FFNN) could predict the cellular data traffic more accurately than the statistical ARIMA model \cite{b16}. While C. Zhang et al. predicted the citywide traffic using Convolutional Neural Network (CNN) by treating data traffic as images \cite{b17}. However, the approaches mentioned above for traffic prediction did not explain the performance of their system for the scenario where there is a sudden change in traffic pattern compared to daily patterns due to situations like pandemics or summer holidays where the traffic pattern would be entirely different in many areas compared to regular days. In this paper, we propose a model for traffic prediction that can adapt to any situation in less time.
The layout of the remaining paper is as follows. Section II delineates the System's Model. Section III contains an analysis of our experiments and results. Section IV illustrates the conclusions and discusses future areas of exploration.

\section{SYSTEM MODEL}\label{AA}
Our system model mainly comprises of two parts: Cellular Traffic Predictor (CTP) and Premeditated Cell Transformer (PCT). The main aim of CTP is to predict the traffic and the priorities that would be generated the next day from a given area when trained with the previous traffic and priority data. While PCT trades the coverage area between every neighboring Base Station (BS) pair until the load is divided equally among all the Bases stations, thereby predicting the area that should be covered by each BS the next day.
Generally, while designing a cellular network, the geographical area is divided into small hexagonal regions called cells, and at the center of hexagons,BSs are installed. 
Hexagonal cells are preferred instead of circular cells since in hexagonal cells, frequency reuse is feasible, and they can cover the entire area without overlapping.  
Each of these cells is divided into 24 microcells in a triangular shape, as shown in Fig.1. The circles in the figure depict the maximum area that can be covered by the BS's. Massive MIMO technology is used since these antenna groups can have the directivity up to \(360^0\) with better throughput and spectrum efficiency. The microcells which are inside the circle and are exterior to the hexagon area of a BS are called the Scope-1 microcells with respect to the corresponding BS. Whereas Scope-0 microcells w.r.t a BS are those microcells which are inside the hexagonal area covered by that BS and are at the border to the neighboring hexagons. Hence after predicting the traffic of each of these microcells, to balance the load, neighboring BSs exchange some of their Scope-1 and Scope-0 microcells until both carry the equal load.

\begin{figure}
  \centering
  \includegraphics[width=0.5\textwidth]{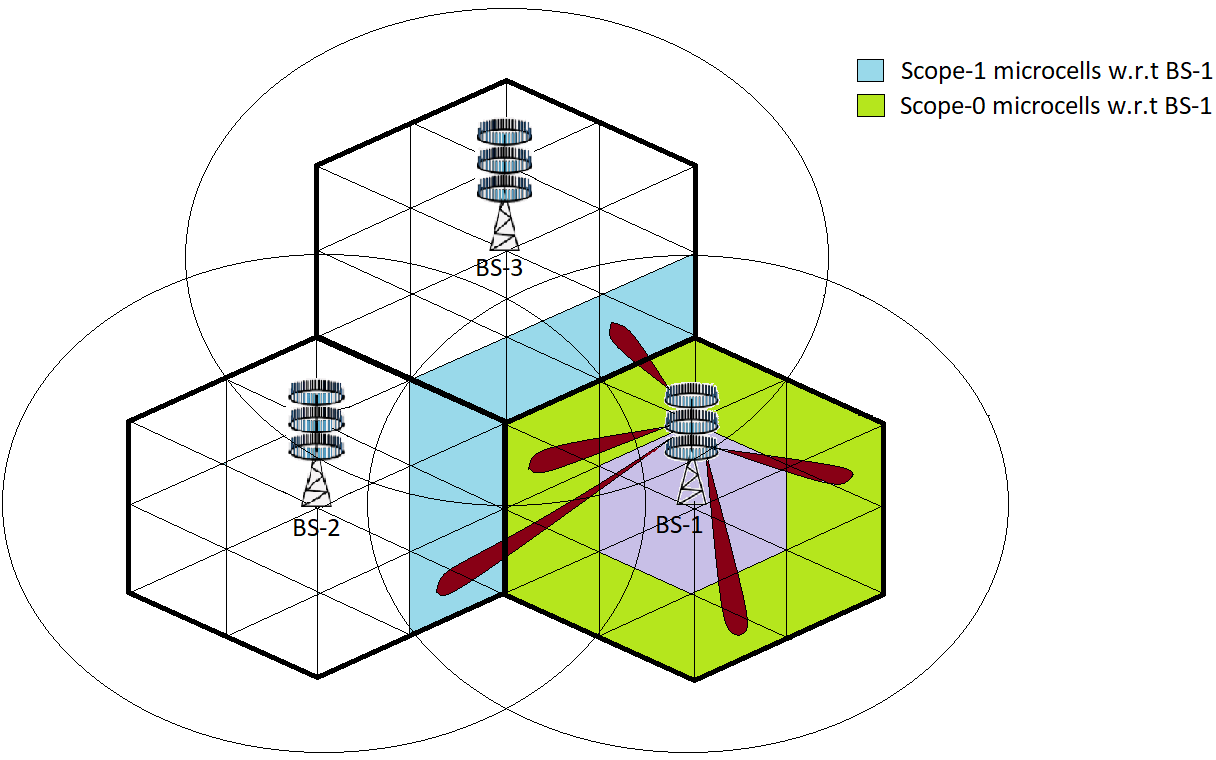}
  \caption{ Wireless cellular network having 3 Cells with 24 microcells in each}
\end{figure}

\subsection{Architecture and Hypothesis}

\begin{figure*}
  \centering
  \includegraphics[width=\textwidth]{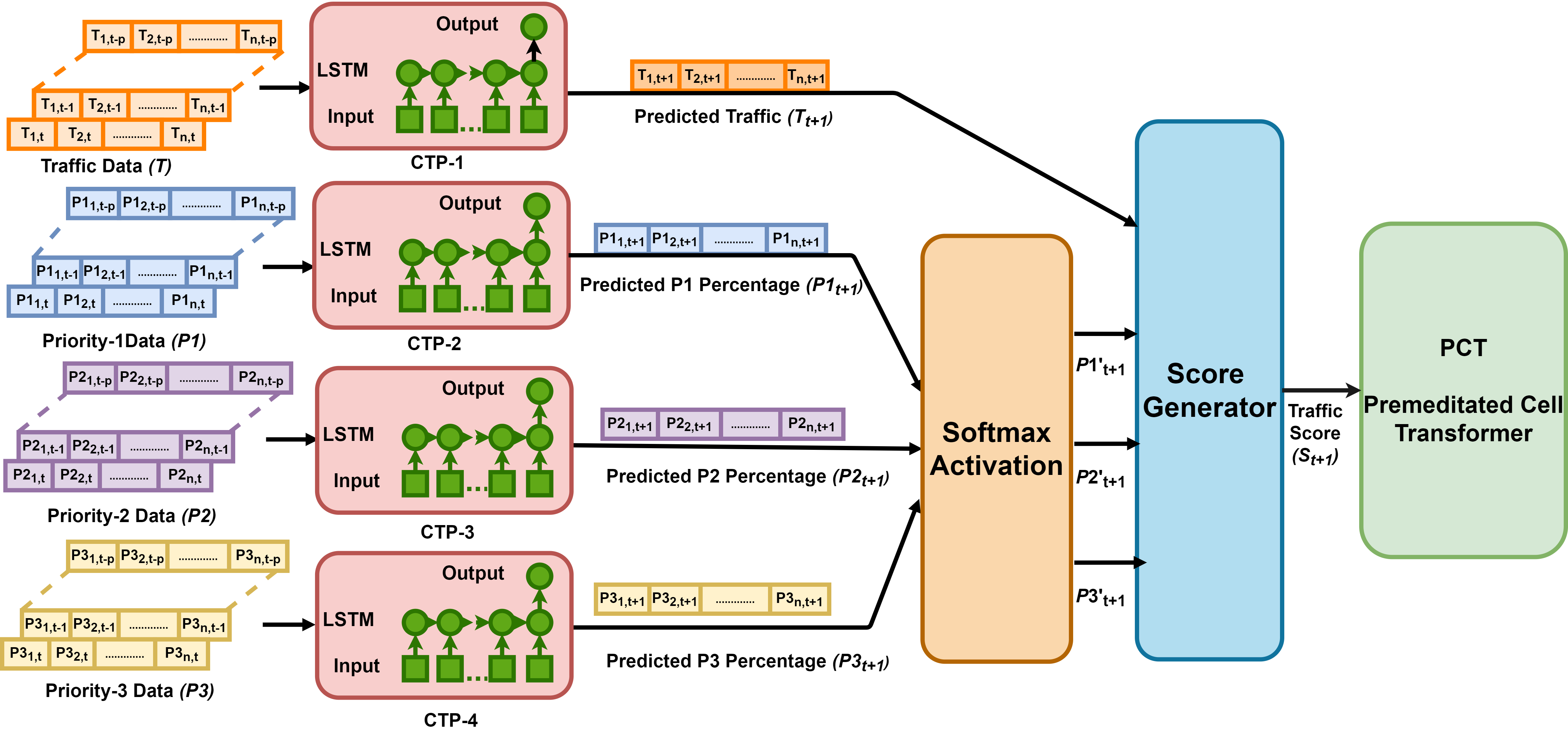}
  \caption{Architecture of System Model.}
\end{figure*}

The architecture of the system model is shown in Fig.2. Consider data vectors $T_t$,$T_{t-1}$,...$T_{t-p}$ containing the time-series data traffic of each microcell at timesteps $t$,$t$-1,...$t$-$p$ as given in Eq.(1-3)

\begin{equation}
T_t = \{T_{1,t}, T_{2,t}...,T_{n,t}\}
\end{equation}

\begin{equation}
T_{t-1} = \{T_{1,t-1}, T_{2,t-1}...,T_{n,t-1}\}
\end{equation}
\begin{equation}
T_{t-p} = \{T_{1,t-p}, T_{2,t-p}...,T_{n,t-p}\} 
\end{equation}
where $(p+1)$, $n$ represents the sequence length and the total number of microcells respectively. Total number of BSs \(N=\frac{n}{24}\).
The data traffic of every microcell across all timesteps is represented as $T$.In general,$T_{i,j}$ represents the traffic value at $i^{th}$ microcell at $j^{th}$ timestep.
Similarly, $P1_{i,j}$, $P2_{i,j}$, $P3_{i,j}$ represents the percentage of traffic at  Priority-1, Priority-2, Priority-3 respectively at $i^{th}$ microcell at $j^{th}$ timestep, and the traffic, priorities data across all timestamps are represented as $T,P1,P2,P3$ and are given in Eq.(4-7).  
\begin{equation}
T = \{T_{t}, T_{t-1}...,T_{t-p}\}
\end{equation}
\begin{equation}
P1 = \{P1_{t}, P1_{t-1}...,P1_{t-p}\}
\end{equation}
\begin{equation}
P2 = \{P2_{t}, P2_{t-1}...,P2_{t-p}\}
\end{equation}
\begin{equation}
P3 = \{P3_{t}, P3_{t-1}...,P3_{t-p}\}
\end{equation}
Since a given microcell at any given timestep, the sum of all the priorities is equal to 100\%, Eq.(8) should be satisfied.
\begin{equation}
P1_{i,j}+P2_{i,j}+P3_{i,j}=1
\end{equation}
The CTP predicts the next timestep traffic and priority data of each microcell by taking in their previous data and is presented in Eq.(9-12).
\begin{equation}
T_{t+1}=CTP1(\{T_{t}, T_{t-1}...,T_{t-p}\})
\end{equation}
\begin{equation}
P1_{t+1}=CTP2(\{P1_{t}, P1_{t-1}...,P1_{t-p}\})
\end{equation}
\begin{equation}
P2_{t+1}=CTP3(\{P2_{t}, P2_{t-1}...,P2_{t-p}\})
\end{equation}
\begin{equation}
P3_{t+1}=CTP4(\{P3_{t}, P3_{t-1}...,P3_{t-p}\})
\end{equation}
where $CTP1, CTP2, CTP3, CTP4$ are the functions that fit the LSTM deep neural networks. Since the predicted priorities do not always satisfy Eq.(8), it can be presumed that they satisfy Eq.(13),
\begin{equation}
P1_{i,t+1}+P2_{i,t+1}+P3_{i,t+1} \neq 1
\end{equation}

To make them satisfy Eq.(8), they are activated using softmax function and the corresponding activated outputs $P1'_{i,t+1}$, $P2'_{i,t+1}$, $P3'_{i,t+1}$ presented in Eq.(14-16),

\begin{equation}
P1'_{i,t+1}= \frac{\exp{P1_{i,t+1}}}{\exp{P1_{i,t+1}}+\exp{P2_{i,t+1}}+\exp{P3_{i,t+1}}}
\end{equation}

\begin{equation}
P2'_{i,t+1}= \frac{\exp{P2_{i,t+1}}}{\exp{P1_{i,t+1}}+\exp{P2_{i,t+1}}+\exp{P3_{i,t+1}}}
\end{equation}

\begin{equation}
P3'_{i,t+1}= \frac{\exp{P3_{i,t+1}}}{\exp{P1_{i,t+1}}+\exp{P2_{i,t+1}}+\exp{P3_{i,t+1}}}
\end{equation}

It is clear from Eq.(3, 4, and 5) that, activated outputs satisfy Eq.(17).
\begin{equation}
P1'_{i,t+1}+P2'_{i,t+1}+P3'_{i,t+1}=1
\end{equation}

Generally, the uplink bandwidth allotted for P1 is four times the bandwidth allotted for P3 and two times the bandwidth allotted for P2\cite{b18}. Hence it is considered that P1 is four times more important than P3 and two times more important than P2 for the following analysis. Since the analysis of distributing the load among all BSs by considering traffic, P1, P2, P3 data of each microcell is severe; a new parameter called traffic score($S$) is introduced that depends on each of the traffic, P1, P2, P3 data. The traffic score for for $i^{th}$ cell and for all cells are given in Eq.(18),(19) respectively.
\begin{equation}
S_{i,t+1}=4*T_{i,t+1}*P1'_{i,t+1}+2*T_{i,t+1}*P2'_{i,t+1}+T_{i,t+1}*P3'_{i,t+1}
\end{equation}
\begin{equation}
S_{t+1} = \{S_{1,t+1}, S_{2,t+1},...,S_{n,t+1}\}
\end{equation}
Based on the traffic scores($S_{t+1}$) and the location data of microcells, load balancing is done by PCT.

\subsection{Cellular Traffic Predictor}
The CTP consists of a multistage LSTM deep neural network capable of predicting data traffic and priorities data based on previous data. LSTMs\cite{b19} are a special kind of Recurrent Neural Network (RNN), capable of remembering information for long periods. Each LSTM layer has a chain-like structure with the repeating modules called LSTM-cells.  The structure of LSTM-cell is shown in Fig.3. Here, $I_t$ is the input to the current cell, $y_t,y_{t-1}$ denotes current and previous hidden states and $C_t,C_{t-1}$ denotes the current and previous cell states. There are three different gates in an LSTM-cell that control the flow of information. They are: forget gate ($\Gamma_t$), input gate ($i_t$), and output gate ($\Omega_t$). The forget gate removes the information, which is less important and keeps only the information that is required for understanding things. The input gate is responsible for adding information to the cell state while the output gate decides what the next hidden state should be. 
\begin{figure}[h!]
  \centering
  \includegraphics[width=0.5\textwidth]{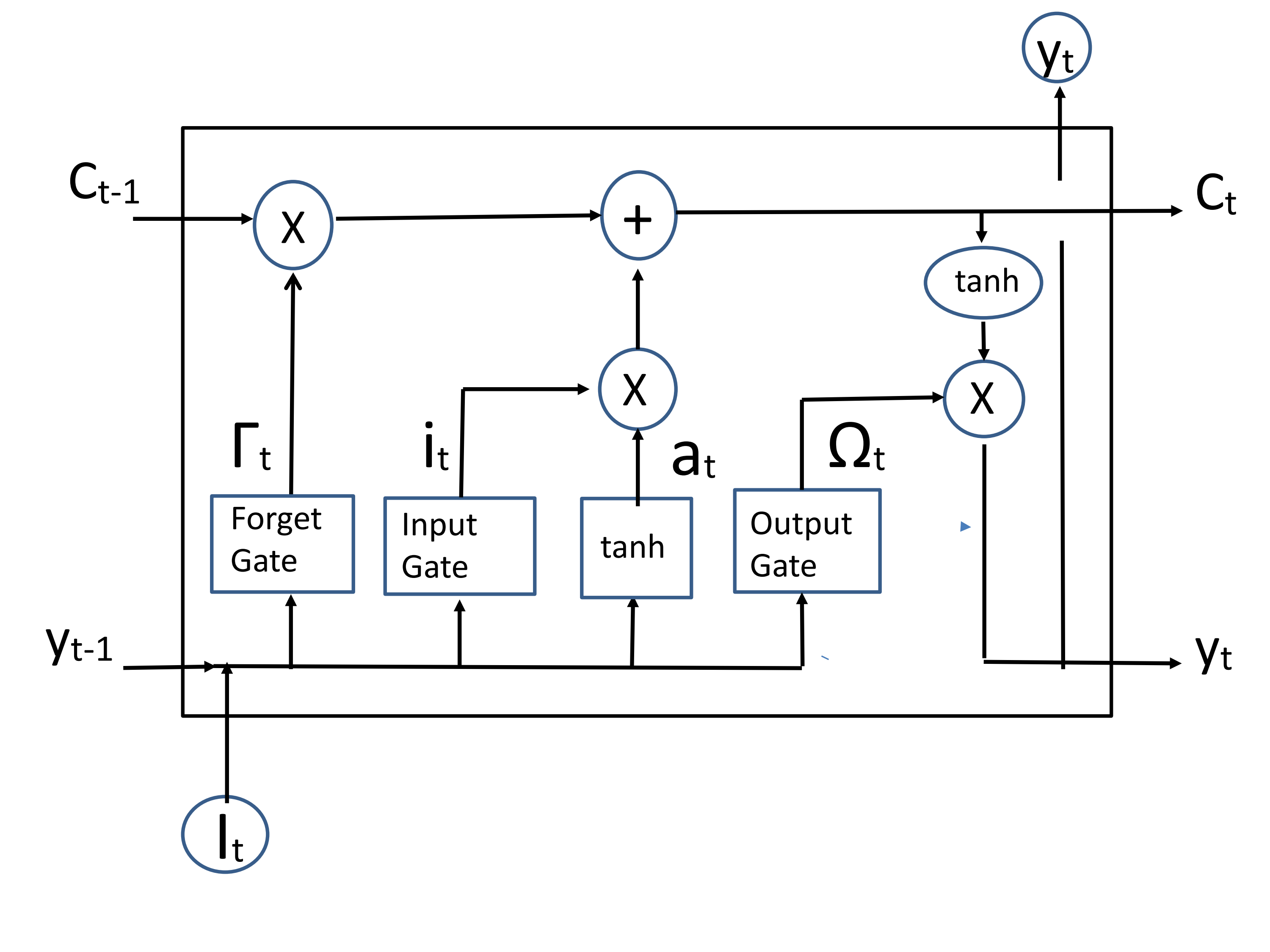}
  \caption{LSTM cell. \\
  Source:\cite{b20}}
\end{figure}

The expressions for all the gates, hidden state, cell state are as given in Eq.(20-25),

\begin{equation}
\Gamma_t=Sig(U_f[y_{t-1},I_t]+\phi_f)
\end{equation}
\begin{equation}
i_t=Sig(U_i[y_{t-1},I_t]+\phi_i)
\end{equation}
\begin{equation}
a_t=tanh(U_a[y_{t-1},I_t]+\phi_a)
\end{equation}
\begin{equation}
C_t=\Gamma_t*C_{t-1}+i_t*a_t
\end{equation}
\begin{equation}
\Omega_t=Sig(U_o[y_{t-1},I_t]+\phi_o)
\end{equation}
\begin{equation}
y_t=\Omega_t*tanh(C_t)
\end{equation}

\begin{figure}[h!]
   
  \centering
  \includegraphics[width=0.5\textwidth]{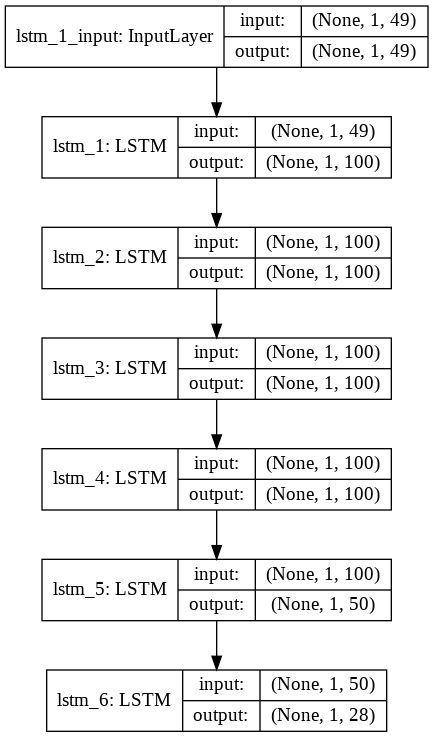}
  \caption{Deep Neural Network architecture used in CTP.}
\end{figure}

Where $Sig$ represents the sigmoid function which is given by \(Sig(x)=\frac{1}{1+\exp(-x)}\) and $*$ denotes the element-wise multiplication. While $U_(.)$ and $\phi_(.)$ denote the vectors of weights and biases corresponding to the respective gates, input, hidden layer. The architecture of the LSTM deep neural network used for this work is shown in Fig.4.

\subsection{Premeditated Cell Transformer}

\begin{figure*}
   
  \centering
  \includegraphics[width=\textwidth]{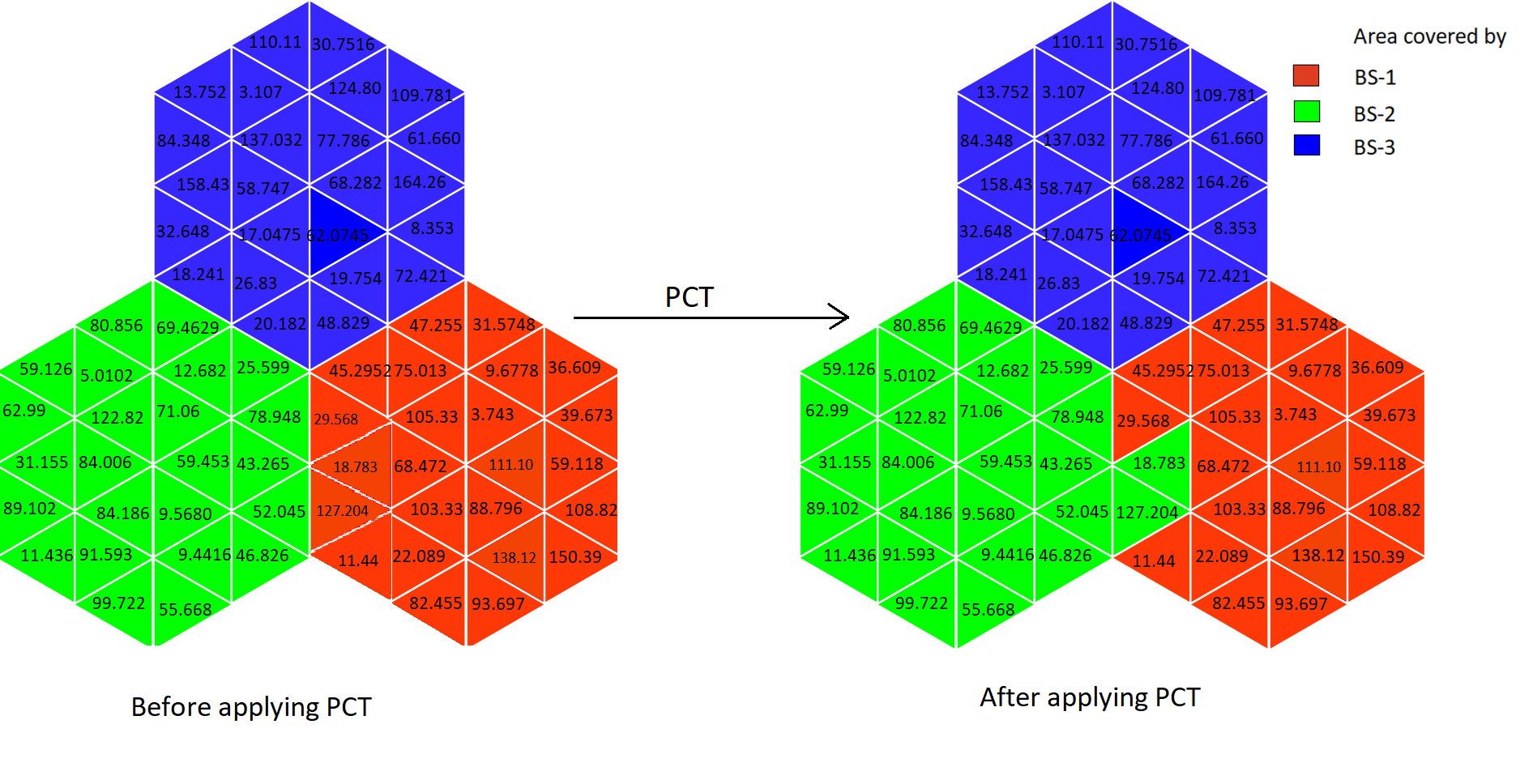}
  \caption{Wireless Network Layout-1}
\end{figure*}

\begin{figure*}[h!]
   
  \centering
  \includegraphics[width=\textwidth]{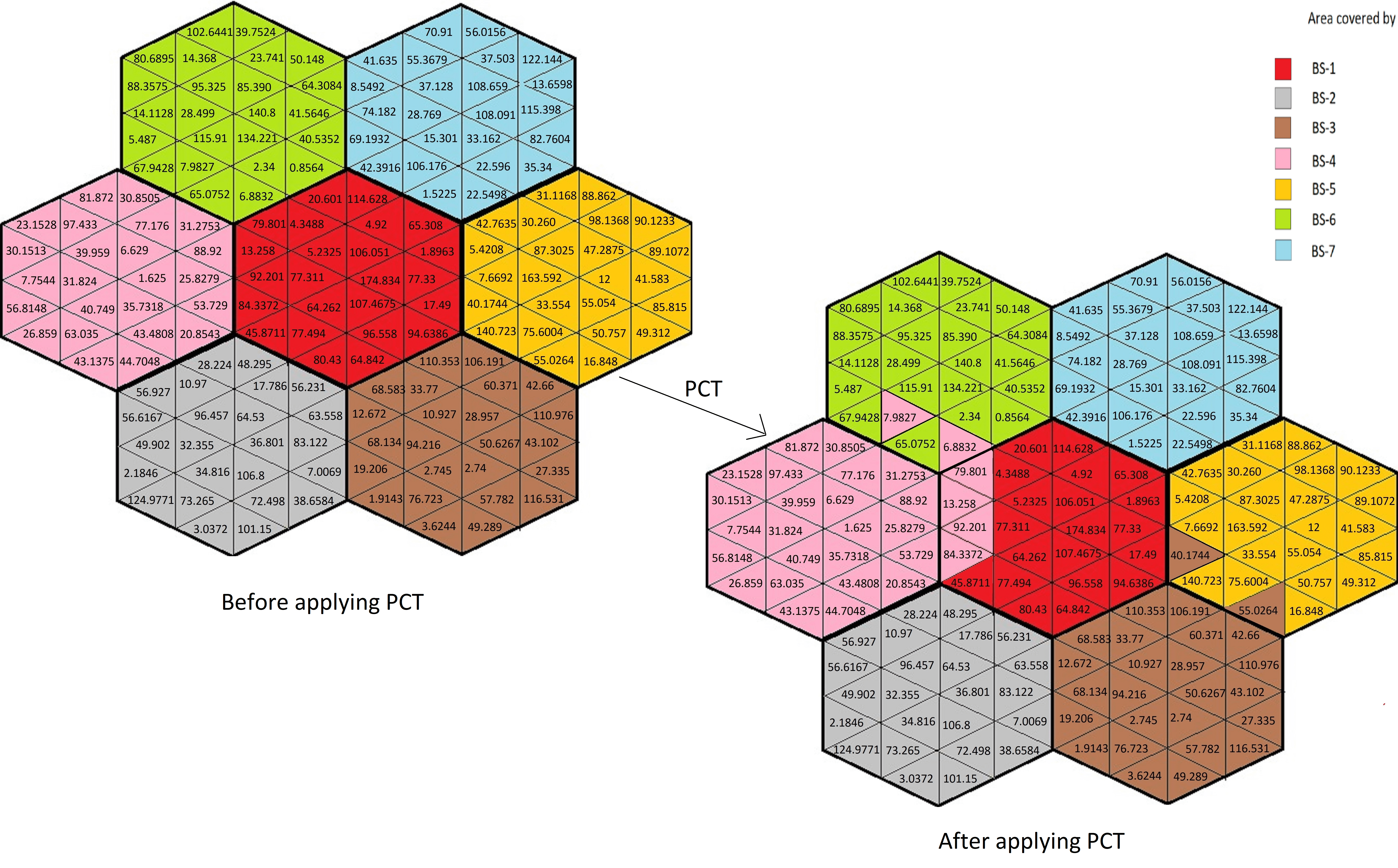}
  \caption{Wireless Network Layout-2}
\end{figure*}

\begin{algorithm}
\scriptsize
\caption{Premeditated Cell Transformer}
${S}$: Set of traffic scores of all triangles where \textbf{S[i]} represents the traffic score of \([quotient(i\div24)]^{th}\) hexagon's \([remainder({i\div24})]^{th}\) triangle; \leavevmode\\
${N}$: Number of Hexagons/BSs;\leavevmode\\
${HL}$: Hexagon's location data;\leavevmode\\
avg\_score=0;\leavevmode\\
iterations=0;\leavevmode\\
\SetKwFunction{FMain}{GetNearByHexagons}
    \SetKwProg{Fn}{Function}{:}{}
    \Fn{\FMain{$num$}}{
        hexagon = [];    //saves BS numbers \leavevmode\\
        \For{$i\gets1$ \KwTo $N$}
        {
            \If{$euclidian\ distance\ ${(HL[i][0], HL[num][0])}\ =$\ 4*height\ of\ triangle\ \textbf{and}\ i\ \neq $num}{
                Add to hexagon;
                }
        }
        \KwRet hexagon;
        }
\SetKwFunction{FMain}{rec}
    \SetKwProg{Fn}{Function}{:}{}
    \Fn{\FMain{addable, min\_index, u, added, HexTraffic}}{
        i,j = addable(u);\leavevmode\\
        \If{$u+1 < length\ of\ addable$}{
        \KwRet HexTraffic,Added;
        }
        \If{$HexTraffic[min\_index]+S[(i-1)*24+j] > avg\_score\ \textbf{or}\ HexTraffic[i]+S[(i-1)*24+j] < avg\_score$}{
        \KwRet rec($addable$, $min\_index$, $u+1$, $added$, $HexTraffic$);
        }
        HexTraffic1\ ,\ Added1\ =\ rec(addable, min\_index, u+1, added, HexTraffic);\leavevmode\\  
        HexTraffic[min\_index]\ =\ HexTraffic[min\_index]\ +\ S[(i-1)*24+j];\leavevmode\\
	    HexTraffic[i]=HexTraffic[i]-S[(i-1)*24+j];\leavevmode\\
	    HexTraffic2,Added2=rec(addable, min\_index, u+1, added+(i,j), HexTraffic);\leavevmode\\
	    \If{$HexTraffic1[min\_index] < HexTraffic2[min\_index]$}{
	    \KwRet HexTraffic2,Added2;
	    }
	        \KwRet HexTraffic1,Added1;
        }
\SetKwFunction{FMain}{GetAddableTriangles}
    \SetKwProg{Fn}{Function}{:}{}
    \Fn{\FMain{num}}{
        addable=[];\leavevmode\\
        nearby=GetNearByHexagons(num);\leavevmode\\
        \For{$each\ i\ \epsilon\ nearby$}{
            \If{jth triangle of ith hexagon falls under scope 1 of num}{
            add (i,j) to the addable list;
        }}
    \KwRet addable;
        }
\SetKwFunction{FMain}{TrafficSharingAlgo}
  \SetKwProg{Fn}{Function}{:}{} 
  \Fn{\FMain{$S$, $N$, $HL$}}{
  HexTraffic, Added be two new arrays.\\
          \For{$i\gets1$ \KwTo $N$}{
    HexTraffic[i] = sum of values from $S$[(i-1)*24+1] to $S$[(i-1)*24+24];
    }
    n=length of S;\\
    avg\_score=\(\sum_{i=1}^{n} \frac{S[i]}{N}\);\leavevmode\\
    L = [1,2,...N] be\ a\ new\ array\ mapped\ to\ array\ HexTraffic such that L[k] maps to HexTraffic[k];\leavevmode\\
    sort elements of L in increasing order of their corresponding mappings in HexTraffic ;\leavevmode\\
    \For{$each\ j\ \epsilon\ L$}{
        addable = GetAddableTriangles(j);\leavevmode\\
		remove the  elements from addable that are present in Added array ;\leavevmode\\
		HexTraffic,NewAdded = rec(addable,j,0,[],HexTraffic); \leavevmode\\
		Added = Added+NewAdded; \leavevmode\\
		iterations=iterations+1;\\}}
\end{algorithm}

After predicting the traffic score values of all the microcells using the LSTM network, the total score is distributed equally among all the BSs using PCT. The basic principle involved in PCT is, the exchange of Scope-1 and Scope-0 microcells between adjacent hexagons until all the hexagons carry nearly equal traffic scores. In order to distribute the total traffic score effectively following steps are used,

\textit{ Step 1}. Find the traffic score under each hexagon (BS).
            
\textit{Step 2}. Compute the average traffic score based on number of hexagons $N$ which is presented in Eq.(26).
\begin{equation}
Average\ Traffic\ Score =  \frac{\sum_{i=1}^{n} S_{i,t+1}}{N}
\end{equation}

\textit{Step 3}. Store the hexagons which have traffic score less than the average score in a list.
 
\textit{Step 4}. Take a hexagon from the obtained list in Step3, which has the least traffic and find out all the scope1 triangles for this hexagon.
\textit{Step 5}. Try all possible selections of triangles obtained from \textit{Step 4} and add the set of triangles with maximum possible traffic to the hexagon, such that the traffic score after adding does not exceed the average traffic score and traffic score of the base tower from which we take does not become less than average traffic score. Discard this hexagon from the list so that it does not involve in further steps.
\textit{Step 6}. Repeat \textit{Step 4} and \textit{Step 5} until the list obtained from \textit{Step 3} becomes empty.
\begin{figure}[ht!]
    \centering
    \begin{subfigure}{.5\linewidth}
        \includegraphics[scale=0.5]{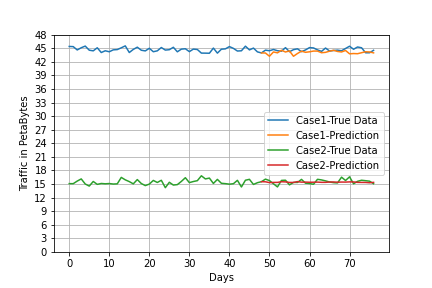}
        \caption{Scenario1}
    \end{subfigure}
    \begin{subfigure}{.5\linewidth}
        \includegraphics[scale=0.5]{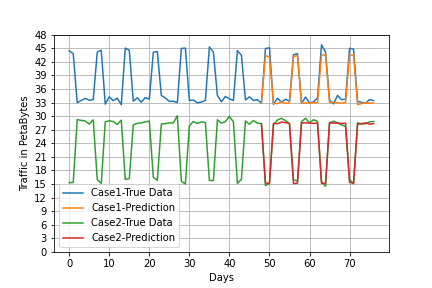}
        \caption{Scenario2}
    \end{subfigure}
    \begin{subfigure}{.5\linewidth}
        \includegraphics[scale=0.5]{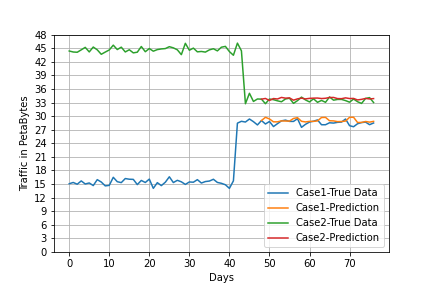}
        \caption{Scenario3}
    \end{subfigure}
    \caption{LSTM prediction results}
\end{figure}
The Algorithm 1 describes the implementation of Premeditated Cell Transformer.
Here all the hexagons are labelled with numbers starting from 1 to N. $S$ contains the traffic scores of all the triangles present in the hexagons at time ${t+1}$. Each hexagon contains 24 triangles numbered from 1 to 24. The traffic of ${j^{th}}$ triangle of ${i^{th}}$ hexagon is given by ${S[i*24+j]}$. All the location co-ordinates of hexagon are stored in ${HL}$, where ${HL[i]}$ represents all co-ordinates corresponding to \textit{ith} hexagon .${HL[i][0}$ contains the ${i^{th}}$ hexagon center location co-ordinates and ${HL[i][j]}$ contains ${j^{th}}$ triangle's location co-ordinates of ${i^{th}}$ hexagon. The  function $rec$ in the algorithm 1, does backtracking to find out the best possible selection of triangles, updates the traffic score values and assigns areas(triangles) to the selected low traffic score tower. The term $iterations$ in the algorithm 1 indicates the number of hexagons that have reached saturation and will not involve in traffic sharing in the further steps.

The analysis of PCT is performed by considering two wireless network layout (WNL) examples. Fig.5 consists of three BSs and all their triangles. Each triangle area's traffic score is also shown. The triangles having the same color get served by the same base tower. The image shows that, after applying the algorithm, some of the traffic (Triangle area) from base towers having higher traffic scores, are taken by base towers with lower traffic scores. Before optimization, the approximate traffic scores under the base towers are 1529.24, 1356.02, 1661.87, and after optimization, the values are 1529.24, 1502.008, 1515.8195 thereby decreasing the variance of all the traffic scores making the network optimized. Fig.6 consists of seven base towers and all their triangles. Each triangle area's traffic score is also shown. The triangles having the same color get served by the same base tower. The image shows us that after applying the algorithm, few of the traffic (Triangle area) from base towers having a higher traffic score is taken by base towers having lower traffic scores. Before optimization, the approximate traffic scores under the base towers are 1571.1097, 1266.1715, 1199.427, 1003.5461, 1438.0893, 1316.9314,1309.004 and after optimization, the values are 1301.5125, 1266.1715, 1294.6283, 1288.0092, 1342.8885, 1302.0655, 1309.0043 thereby decreasing the variance of all the traffic scores making the network optimized.

\section{EXPERIMENTS AND RESULTS}

\begin{figure}
    \centering
  \includegraphics[width=0.5\textwidth]{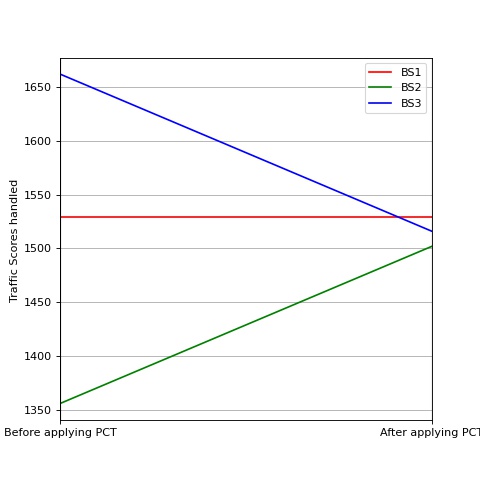}
  \caption{Traffic scores handled by BSs in WNL-1}
\end{figure}

\begin{figure}
   
  \centering
  \includegraphics[width=0.5\textwidth]{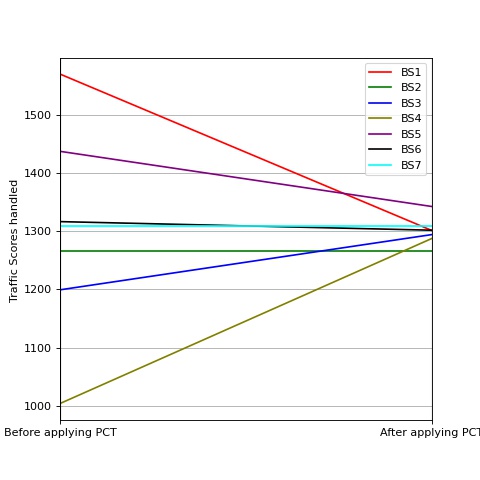}
  \caption{Traffic scores handled by BSs in WNL-2}
\end{figure}
           
Since there is no dataset available online containing all traffic, P1, P2, P3 data in one place, we prepared our dataset. The proposed architecture is trained and tested through a random dataset generated using a Gaussian distribution\cite{b21}. For each triangle, after considering some random base values, for each of the traffic, P1, P2, P3, random values generated using Gaussian distribution with mean as 0 and standard deviation as one are added to this base value generate the whole dataset. The probability density for a gaussian distribution is given in Eq.(27).
\begin{equation}
p(x)=\frac{1}{\sqrt{2\pi{\sigma}^2}}\exp{-\frac{(x-\mu)^2}{2{\sigma}^2}}
\end{equation}

Where \(\sigma\), \(\mu\) represents the standard deviation and the mean, respectively.
Evaluating the performance of the model is tested with three scenarios to check how well it can adapt to situations. In scenario-1, it is considered that the data traffic does not show much variation between weekdays and weekends. While in scenario-2, the data traffic shows variation between weekdays and weekends. In scenario-3, the traffic may change due to sudden changes like vacations, pandemic situations, etc. Scenario-3 is again classified into two classes, where in the first, the sudden change of traffic happens in the distant past, while in the second, the sudden change of traffic happens in the recent past.
\begin{figure}
   \centering  
  \includegraphics[width=0.5\textwidth]{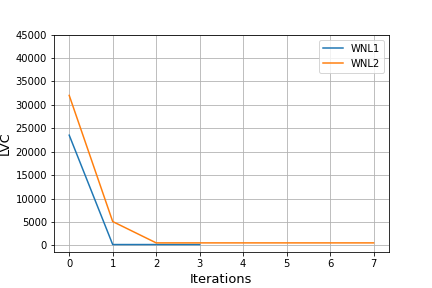}
  \caption{Variation of LVC with iterations}
\end{figure}

\begin{figure}
  \centering   
\includegraphics[width=0.5\textwidth]{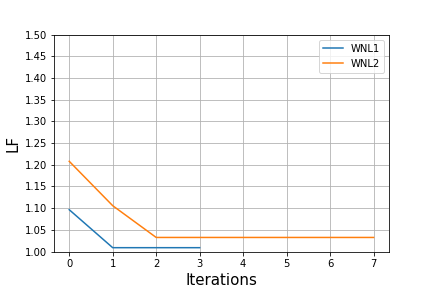}
  \caption{Variation of LF with iterations}
\end{figure}
Furthermore, the dataset contains 2800 training samples for each of the traffic, P1, P2, P3. It comprises 700 training samples from each scenario-1 and scenario-2 and the remaining 1400 from scenario-3. While each test set contains 77 samples, each representing data traffic for each day, of which the model predicts the traffic from the 50th sample to the 77th sample. The prediction results for data traffic are shown in Fig.7. Scenario-2 is explained with two cases where in the first case, the data traffic is high on weekends compared to weekdays, and in the second case, the data traffic is low on weekends compared to weekdays. Similarly, for scenario-3, in the first case, data traffic decreases due to sudden changes, while in second, the data traffic increases due to sudden changes. It can be observed from Fig.7(c) that the model can adapt to the new situations very well within two to three days. The same model is used to predict the P1, P2, P3 data also.

Furthermore, Fig.8 and 9 show the traffic scores handled by each BS before and after applying PCT, considering the WNL shown in Fig.5 and Fig.6, respectively. To study the QoS and smoothness of the load distribution in the network, we define two parameters: The load variance coefficient (LVC) and load factor (LF) and they are presented in Eq.(28),(29) respectively. LVC indicates the deviation in all the BSs' traffic scores from the average traffic score, whereas LF is the ratio of the average traffic score to the peak traffic score.   
\begin{equation}
LVC=\frac{\sum_{j=1}^{N} (\sum_{i=24*(j-1)+1}^{24*(j-1)+24} S_{i,t+1} - \frac{\sum_{i=1}^{n} S_{i,t+1}}{N})^2}{N-1}
\end{equation}
\begin{equation}
LF= \frac{AverageTrafficScore\ of\ a\ cell}{PeakTrafficScore\ of\ a\ cell}
\end{equation}

The decrease in LVC and LF indicates that the traffic is distributed almost evenly among all BSs, thereby increasing the effective bandwidth and the QoS. The variation of LVC and LF for WNL1 and WNL2 are presented in Fig.10 and 11, which implies that there is a gradual decrease in LVC and LF as the $iterartions$ increases, thus improving the QoS. For WNL1, LVC decreases from 32013 to 540.2, and for WNL2, it decreases from 23524.6 to 185.5 while LF decreases from 1.2 to 1.03 for WNL1 and from 1.1 to 1.008 for WNL2. 

\section{CONCLUSION}

A novel load balancing scheme for 6G is proposed through which the traffic load is divided almost equally among all the BSs. It is based on two algorithms, CTP and PCT. CTP predicts the traffic, P1, P2, P3 data using the LSTM network while PCT distributes the total traffic score among all BSs. Distribution of load is done in such a way that the LVC shows a decrease of 98.31\% and 99.21\% for WNL1, WNL2, respectively. Results shown by LSTM and PCT infers that the performance of this method proposed is best or comparable to the existing methods. In the future, the load balancing analysis can be extended by considering the losses in the wireless channel that become significant with the distance.

\end{document}